\documentclass[aps,prl,superscriptaddress,twocolumn,showpacs]{revtex4}
\usepackage{graphics}
\usepackage{epsfig}
\usepackage{color}
\usepackage{stackrel}
\usepackage{amsfonts}
\usepackage{wrapfig}
\usepackage{pstricks}
\usepackage{pst-node}
\usepackage{bm}
\usepackage{dcolumn}
\usepackage{multirow}
\usepackage{epstopdf}
\usepackage{subfigure}
\usepackage{amssymb}
\usepackage{amsmath}
\usepackage{commath}
\usepackage{graphicx,bm}
\usepackage{verbatim}
\usepackage{booktabs}
\usepackage[para]{threeparttable}
\usepackage{etoolbox}
\usepackage{lipsum}
\usepackage{marvosym}
\usepackage{wasysym}

\begin{document}
\title{Excitons in periodic potentials}

\author{Dinh~Van~Tuan}
\altaffiliation{vdinh@ur.rochester.edu}
\affiliation{Department of Electrical and Computer Engineering, University of Rochester, Rochester, New York 14627, USA}

\author{Hanan~Dery}
\altaffiliation{hanan.dery@rochester.edu}
\affiliation{Department of Electrical and Computer Engineering, University of Rochester, Rochester, New York 14627, USA}
\affiliation{Department of Physics and Astronomy, University of Rochester, Rochester, New York 14627, USA}

\begin{abstract}
The energy band structure of excitons is studied in periodic potentials produced by the short-range interaction between the exciton and electrons of Wigner or Moir\'{e} lattices. Treating the exciton as a point-like dipole that interacts with the periodic potential, we can solve a simple one-body problem that provides valuable information on excitons in many-body problem settings. By employing group theory, we identify the excitonic energy bands that can couple to light and then quantify their energy shifts in response to a change in the period of the potential. This approach allows us to emulate the response of optically active exciton and trion states to a change in electron density. We gain important insights on the relation between the electron order in a Wigner crystal and the energy blueshift of the bright exciton. We discuss the consequences of this relation in the context of optical absorption experiments in monolayer semiconductors. 
\end{abstract}
\maketitle

Excitons in semiconductors allow us to explore intriguing many-body phenomena in condensed matter physics \cite{HaugBook,Combescot_Book,Wang_RMP18}. Through changes in energy, fine structure, and oscillator strength of excitons in response to electrostatic gating or magnetic fields, we can learn about Fermi-edge singularities \cite{Mahan_PR67b,Skolnick_PRL87}, composite states \cite{VanTuan_PRL22,VanTuan_PRB22}, proximity effects \cite{Scharf_PRL17,Choi_NatMater22}, Fermi polarons \cite{Sidler_NatPhys17,Ravets_PRL18}, Wigner crystals \cite{Smolenski_Nat21,Zhou_Nat21}, fractional filling of Moir\'{e} valleys \cite{Xu_Nat20,Regan_Nat20}, as well as dynamical screening and exchange-correlation interactions \cite{SchmittRink_PRB86,VanTuan_PRB19}. Many of these problems are hard to tackle from a theoretical standpoint since it is impossible to solve the many-body wave function of the exciton and all interacting electrons. Instead, approximated solutions usually treat the excitonic state as a quasiparticle that is dressed by the interacting electrons \cite{Haug_PQE,Hawrylak_PRB91,Bronold_PRB00,Suris_PSSb01,Esser_PSSb01,VanTuan_PRX17,Efimkin_PRB17, Chang_PRB19,Glazov_JCP20,Rana_PRB20,Imamoglu_CR21}. That is, the exciton creation polarizes the electron gas, and the resulting  displacement of electrons modifies the exciton wave function.  An important question that arises is whether experimental results can be explained without the polarization of the electron gas. 

The motivation for asking this question is that experiments in electrostatically-doped semiconductors are often conducted in the limit $r_x \ll r_s$, where $r_x$ is the exciton radius and $r_s$ is the average distance between two nearby conduction-band electrons \cite{Wang_NanoLett17,Courtade_PRB17,Smolenski_PRL19,Wang_PRX20,Liu_PRL20,Liu_NatComm21,Li_NanoLett22}. Given that electrons are far more influenced by the long-range Coulomb interaction between them than by the short-range interaction with the exciton, displacement of electrons in this system can be energetically costly. We argue in this Letter that as long as $r_x \ll r_s$,  the excitonic absorption spectrum can be well explained by assuming that electrons are `frozen' at fixed positions. The short-range interaction between the exciton and electron is replaced by a static potential, allowing us to readily solve such  exciton problem.

There are two ways to study the interaction between the exciton and the potential produced by a frozen landscape of electrons in the host material. One approach is to assume that electrons are fixed at random positions, and then study the exciton wave function through its phase shift and eventual Anderson localization when the charge density increases \cite{VanTuan_PRB2012}. The second approach is to assume that the electrons are ordered, as in the case of a Wigner crystal \cite{Wigner_PR34}. One can then study the dependence of the exciton wave function on electron density through the ensuing change in lattice constant of the electrons crystal. We take the second approach in this Letter, allowing us to calculate the exciton energies by standard band theory  \cite{Voronov_PSS03,Shimazaki_PRX17}. We employ group theory and test which states couple to light and how order (symmetry) affects the results. Hereafter, the crystal, lattice and Brillouin zone refer to the Wigner crystal and not the semiconductor atomic crystal. 

\begin{figure*}
\includegraphics[width=18cm]{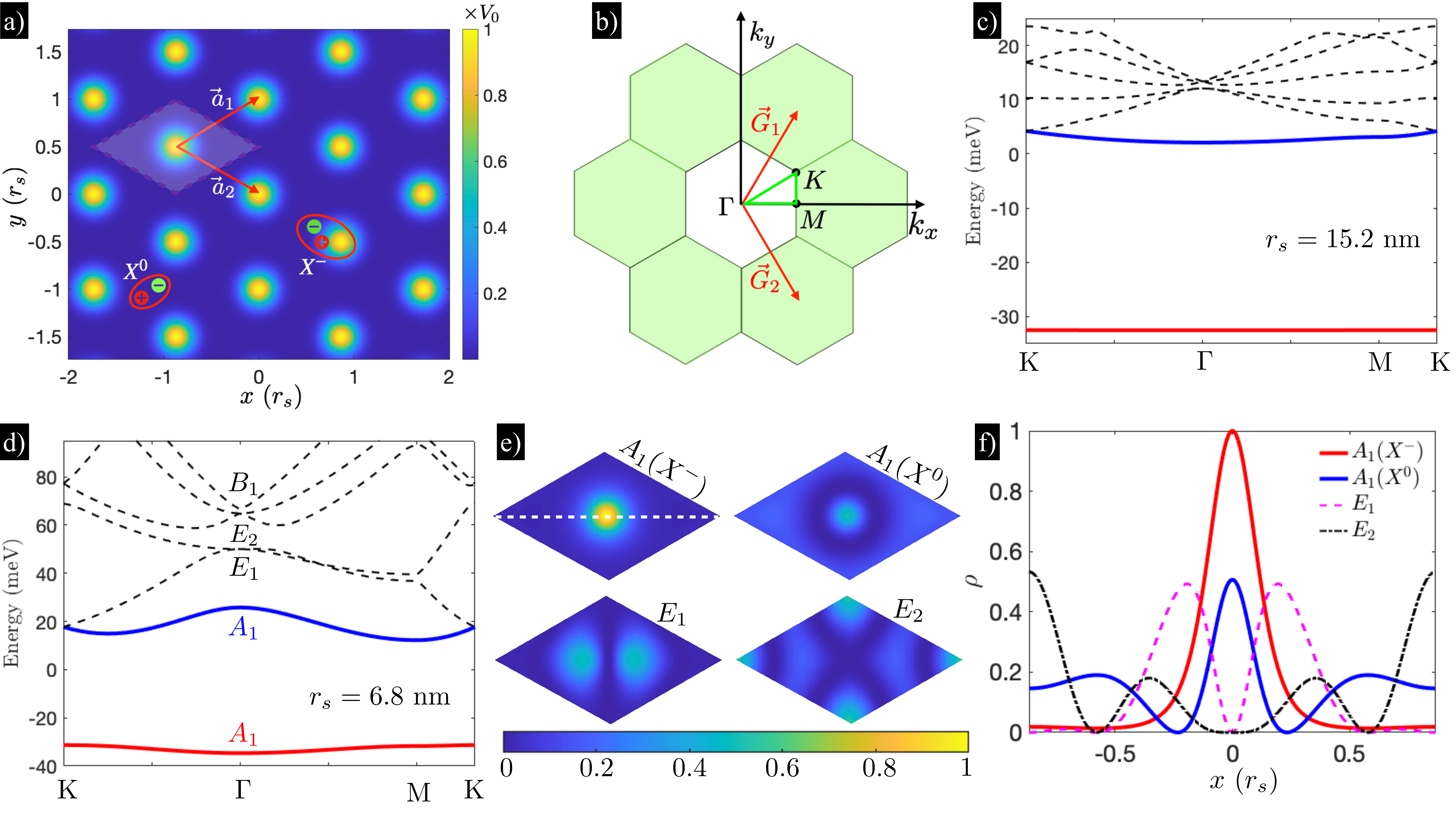}
 \caption{   (a) Illustration of the short-range potential profile experienced by an exciton in a triangular electron lattice. The unit cell and basis vectors ${\bf a}_{1,2}$ are highlighted, where the distance between two neighboring sites is $r_s = |{\bf a}_{1,2}|$. An electron-hole pair propagating in this lattice can form bound or unbound states with the lattice ($X^-$ or $X^0$), corresponding to trion or exciton energy bands, respectively. (b) The first (white) and second (green) Brillouin zones of the triangular lattice, along with the reciprocal-lattice basis vectors  ${\bf G}_{1,2} = 2\pi(1/\sqrt{3},\, \pm 1)/r_s$ and high-symmetry points $\Gamma$, M, and K. (c) and (d) Exciton band structures along axes between high-symmetry points, shown for lattice constants $r_s=15.2$ and 6.8~nm, or equivalently, electron densities $n=5 \times 10^{11}$ cm$^{-2}$ and $n=2.5 \times 10^{12}$ cm$^{-2}$, respectively. The energy bands in (d) are labeled with their $\Gamma$-point IRs (Table \ref{tab:CharacterTabD6}).  (e) Square amplitude colormaps of the $\Gamma$-point wave functions in the unit cell, where (f) shows these results along the axis marked by the dashed white line in (e).} \label{fig:BandStructure} 
\end{figure*}

Our study focuses on two-dimensional (2D) crystals. Given that $r_x \ll r_s$, where $r_s$ is now the lattice constant, we neglect the internal relative motion of the electron and hole components in the exciton and treat the small exciton complex as one body with translational mass $M$. Furthermore, the  lattice sites are assumed to be made of electrons with different quantum numbers (spin and/or valley) than those of the electron in the exciton \cite{footnote,Robert_NatComm21}. This distinguishability not only allows the exciton to bind the lattice and form a trion-like state, but it will be shown to introduce a strong energy blueshift of exciton-like states. Without loss of generality, we consider a periodic Gaussian function to describe the short-range crystal potential exerted on the exciton by the  lattice, 
\begin{equation}
V(\mathbf{r}) = \sum_{n_1, n_2} V_0 \exp\left(- |{\bf r} - n_1 {\bf a}_1 - n_2 {\bf a}_2 |^2/w^2\right)\,.
\label{Eq:Poten}
\end{equation}
$\mathbf{r}$ is the  center-of-mass coordinate of the exciton. $n_{1,2}$ are integers, ${\bf a}_{1,2}$ are the lattice basis vectors, $w$ is the potential range, and $V_0$ is its amplitude. Figures \ref{fig:BandStructure}(a) and (b) show a density plot of the potential in a triangular lattice and the corresponding Brillouin zone, respectively.  

We employ the pseudopotential method to calculate the exciton band structure \cite{Chelikowsky_PRB76}. The Bloch wave function 
\begin{equation}
\psi_{\bf k}({\bf r}) = e^{i\bf k.r} u_{\bf k}({\bf r})=   e^{i\bf k.r} \sum_{\bf G} u_{\bf k}({\bf G}) e^{i\bf G.r},
\end{equation}
is written as a modulated periodic function $u_{\bf k}({\bf r})$, expanded as a sum over reciprocal lattice vectors $\bf G$. The Fourier components $u_{\bf k}({\bf G}) $  and exciton eigenenergies $E_{\bf k}$ are obtained from the  matrix equation
\begin{equation}
  \sum_{\bf G'} H_{\bf GG'} \left( {\bf k} \right) \,\,  u_{\bf k}({\bf G'})  = E_{\bf k}  \ u_{\bf k}({\bf G})\,,  
\end{equation}
where the Hamiltonian in reciprocal space reads
\begin{equation}
H_{\bf GG'} \left( {\bf k} \right)=
 \frac{\hbar^2({\bf k+ G})^2}{2M}   \delta_{\bf G,G'}  +  V({\bf G-G'}). 
\end{equation}
$V({\bf G-G'})$ is the Fourier transform of the crystal potential per unit cell \cite{Cardona_book}. 

Figures~\ref{fig:BandStructure}(c) and (d) show results of the exciton band structure for two lattice constants, $r_s$, related through $nr_s^2 = 2/\sqrt{3}$ to the electron density in a triangular Wigner crystal. The crystal potential parameters are $w=1$~nm and $V_0=-170$~meV, and the exciton translation mass is $M=0.65 m_0$, where $m_0$ is the free electron mass. These are the only three  material-specific parameters, and here we chose them to mimic the behavior in transition-metal dichalcogenide (TMD) monolayers \cite{Yang_PRB22}, where $w$ is comparable to the exciton radius, $V_0$ is chosen to yield a trion-like band  (explained below), and $M$ denotes the center-of-mass rather than reduced mass of the exciton. The zero energy level in our calculations is the exciton resonance energy in the limit $r_s \rightarrow \infty$, corresponding to an intrinsic semiconductor ($n \rightarrow 0$).

The result of having one site per unit cell (Fig.~\ref{fig:BandStructure}(a)), is that only one band has  negative energies, as shown in Figs.~\ref{fig:BandStructure}(c) and (d). The negative-energy band mimics the trion state, which is tightly bound to a lattice site, as shown in Figs.~\ref{fig:BandStructure}(e) and (f) for the square amplitude of the lowest-energy wave function at the $\Gamma$ point ($X^-$). The energy of the trion-like band is $\sim-32$~meV in both Figs.~\ref{fig:BandStructure}(c) and (d), implying that this energy level is weakly dependent on lattice constant  (electron density) and mostly governed by the trapping amplitude of the potential, $V_0$. This behavior persists as long as $w \sim r_x \ll r_s$, leading to strong localization of trion-like states at lattice sites. The nearly flat nature of this energy band implies very large effective mass, or equivalently,  suppressed hopping between neighboring lattice sites. 

The positive energy bands in Figs.~\ref{fig:BandStructure}(c) and (d) describe states in which the exciton tends to stay away from lattice sites (i.e., not bound to the crystal).  Figures~\ref{fig:BandStructure}(e) and (f) show that the $\Gamma$-point wave functions become more extended across the unit cell as the energy of the state increases. To check which of the states can strongly couple to light, we look for wave functions that retain the $1s$ character of the envelope function of the bright exciton  \cite{Voronov_PSS03}.  Before embarking on the symmetry analysis,  we note that focusing on the $\Gamma$-point wave functions is justified when the lattice constant, $r_s$, is much smaller than the photon wavelength needed to create the bright exciton of the semiconductor ($\lambda_x$). Put differently, the exciton momentum following light excitation is much smaller than the width of the Brillouin zone, $ h/\lambda_x \ll 2\pi/r_s$.

\begin{table}
\caption{\label{tab:CharacterTabD6} Character table of point-group $D_6$. The symmetry operators include the identity operation $E$, $\pm 2 \pi/\ell$  rotations around the $z$-axis ($C_\ell(z)$), and $\pi$ rotations around axes that connect nearest-neighbor sites of the triangular lattice ($C'_2$), and next nearest-neighbor sites ($C^"_2$).}
\begin{tabular}{|c||c|c|c|c|c|c|c|}
\hline
\hline
 $\bf D_6$& E &$2C_6(z)$ &$2C_3(z)$&
 $ C_2(z)$ & $ 3C'_2 $ &$3 C^"_2$ &linear functions\\
\hline
$A_1$& +1 & +1 & +1 &+1 & +1 & +1 & - \\
\hline
$A_2$& +1 & +1 & +1 &+1 & -1 & -1 & $z$ \\
\hline
$B_1$& +1 & -1 & +1 &-1 & +1 & -1 & - \\
\hline
$B_2$& +1 & -1 & +1 &-1 & -1 & +1 & - \\
\hline
$E_1$& +2 & +1 & -1 &-2 & 0 & 0 & $ x,y$ \\
\hline
$E_2$& +2 & -1 & -1 &+2 & 0 & 0 & - \\
\hline
\hline
\end{tabular}
\end{table}

Table \ref{tab:CharacterTabD6} shows the character table of point-group $D_6$, corresponding to the $\Gamma$ point of the 2D triangular lattice, where the $z$-axis is along the out-of-plane direction. Among the seven lowest energy states of the $\Gamma$ point that we show in Figs.~\ref{fig:BandStructure}(c) and (d), only the lowest two transform as the identity irreducible representation (IR) $A_1$.  The transformation properties of $A_1$ describe the envelope functions of the ground-state trion or exciton (1$s$), whose coupling to light is strongest. Of the remaining $\Gamma$-point states that are shown in Figs.~\ref{fig:BandStructure}(c)-(f), those that transform like $E_1$ can turn valuable in experiment. The doubly-degenerate IR $E_1$ transforms like an in-plane vector ($x$ and $y$). As such, the selection rule $A_1 \times E_1 = E_1$ mandates that the dipole transition between $E_1$ and $A_1$ states does not vanish, $\left \langle A_1 \left| {\bf p} \right| E_1 \right\rangle \neq 0$, where the momentum operator $\mathbf{p}$ also transforms like an in-plane vector (i.e., represented by $E_1$).  As we show next, the consequences of this property can be used to test if the electrons are ordered in a Wigner lattice.

\begin{figure*}
\includegraphics[width=17cm]{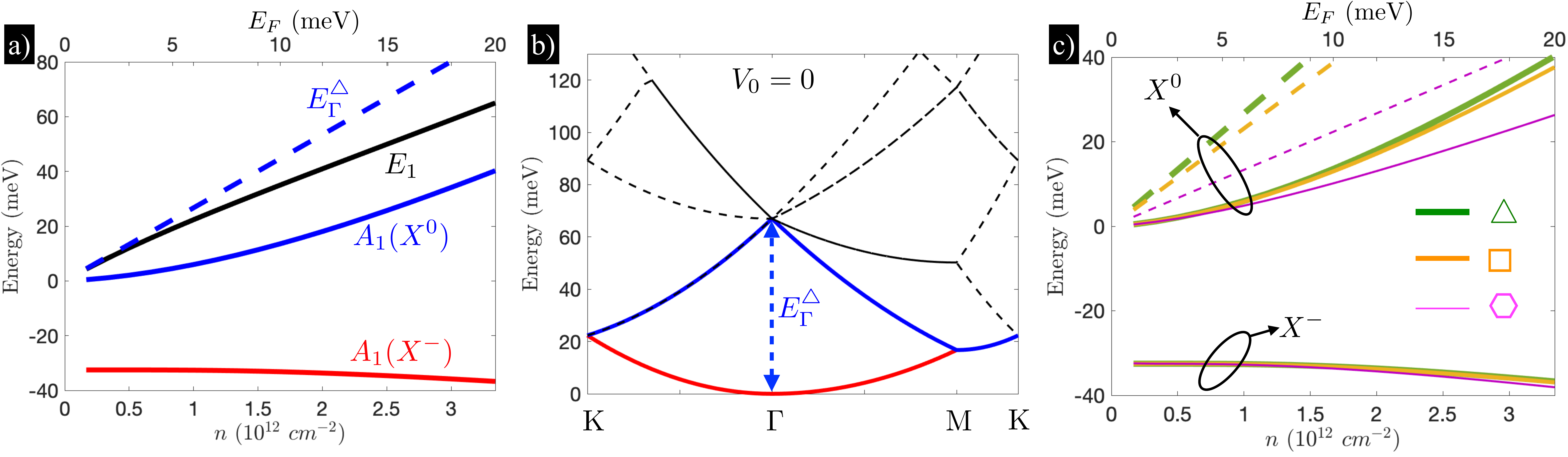}
 \caption{   (a)  $\Gamma$-point energies of the trion $(X^-)$,  bright exciton $(X^0)$, and excited exciton state with $E_1$ symmetry as a function of electron density $n$ (bottom $x$-axis) or Fermi energy $E_F = \hbar^2 \pi n/ m_e$ (top $x$-axis, where $m_e =0.4m_0$). The dashed line denotes the energy $E^\triangle_{\Gamma}$ of the nearly-free exciton model. (b) Energy band structure of the nearly-free exciton model  ($V_0=0$) when $n=2.5 \times 10^{12}$ cm$^{-2}$.  (c) $\Gamma$-point energies of the trion and exciton in three lattice structures: triangular  (green), square  (orange), and honeycomb (magenta).  The dashed lines are the respective $E^\triangle_{\Gamma}$, $E^\square_{\Gamma}$ and $E^{\hexagon}_{\Gamma}$ of the nearly-free exciton model. } \label{fig:TrionExc}
\end{figure*}

Consider a photoexcited semiconductor with trions or excitons in their ground state (1$s$). The center-of-mass envelope functions of these excitonic complexes are described by the two aforementioned $A_1$ states if the electrons in the semiconductor are ordered in a Wigner lattice. If the semiconductor is further excited by a second light source with far-infrared photons, the absorption of these photons is greatly enhanced when their energy matches the energy gap between $A_1$ and $E_1$ states, $E_{E_1}-E_{A_1}$. To verify that the absorption resonance stems from optical transitions of the Wigner-induced exciton band structure, rather than being an accidental occurrence, one can tune the electron density in the semiconductor to control the energy gap $E_{E_1}-E_{A_1}$. The solid lines in Fig.~\ref{fig:TrionExc}(a) show the energy shifts of these states in a triangular Wigner lattice. The energy gaps between $E_1$ (black solid line) and any of the $A_1$ states (red or blue solid lines) increase with electron density. 

Figure~\ref{fig:TrionExc}(a) shows that the energy shift of the trion ($X^-$) has the smallest change when the electron density increases. As long as $w \sim r_x \ll r_s$, the strong localization around lattice sites renders the trion state less susceptible to a change in electron density (Figs.~\ref{fig:BandStructure}(e) and (f)). The slight redshift of the trion energy when the density increases is explained by the slightly improved overlap of the wave function between lattice sites. In sharp contrast, the exciton states are extended, meaning they are susceptible to density changes. Their energy blueshift becomes stronger as the exciton becomes more extended (i.e., higher energy), as can be seen by comparing the smaller blueshift of $A_1(X^0)$ compared with that of $E_1$ in Fig.~\ref{fig:TrionExc}(a). The reason is that the attractive potential from lattice sites tends to lower the energy of the state, and higher-energy states better avoid these sites, as can be seen in Figs.~\ref{fig:BandStructure}(e) and (f). 

To further understand the energy blueshift of extended states, we repeat the calculation by turning off the potential, $V_0 \rightarrow 0$.  The band structure in this nearly-free exciton model is shown in Fig.~\ref{fig:TrionExc}(b). While trion states evolve from the lowest energy band around the $\Gamma$ point, exciton states evolve from the excited energy bands around the $\Gamma$ point. The six-fold state degeneracy of the second state originates from the six $\Gamma$ points at the centers of the second Brillouin zones, highlighted by green hexagons in Fig.~\ref{fig:BandStructure}(b). The energy of this degenerate state is linearly related to electron density in the nearly-free exciton model
\begin{equation}
E^\triangle_{\Gamma} =  \frac{\hbar^2}{M}  \frac{4\pi^2}{\sqrt{3}}n,
\end{equation}
shown by the dashed line in Fig.~\ref{fig:TrionExc}(a).  When $V_0 \neq 0$, the exciton-like states $A_1(X^0)$ and $E_1$ are not completely free, since the lattice potential lowers their energy from that of the nearly-free model. Yet, the linear dependence on charge density is largely kept, especially for higher-energy states as shown in  Fig.~\ref{fig:TrionExc}(a).  The relatively strong energy shift of the exciton compared with that of the trion comes from the difficulty of the exciton state to remain extended when the electron density increases (i.e, smaller distance between lattice sites). 
 
The behavior shown in Fig.~\ref{fig:TrionExc}(a) is corroborated in optical experiments of electrostatically-doped TMD monolayers in which the electrons (or holes) are not necessary ordered    \cite{Wang_NanoLett17,Courtade_PRB17,Smolenski_PRL19,Wang_PRX20,Liu_PRL20,Liu_NatComm21,Li_NanoLett22}. Regardless of such order, the small energy shift of the trion should persist as long as the average distance between two electrons in the interacting electron gas is small compared with the extension of the trion wave function. Similarly, the presence of a Wigner lattice can only enhance the energy blueshift of the exciton but not quench it because the extended exciton states are susceptible to density changes. Before analyzing how order affects the energy shifts, we note that the trion and exciton states in our model have similar attributes to attractive and repulsive polarons in the following sense \cite{Sidler_NatPhys17,Efimkin_PRB17}. The attractive polaron in our case is an exciton that is strongly attracted to lattice sites, whereas the repulsive polaron is an exciton that tends to stay away from them. Both attributes are seen in Figs.~\ref{fig:BandStructure}(e)-(f),  where the lattice site is at the center of the unit cell. Contrary to the Fermi-polaron picture, however, our model explains the energy shifts as a natural consequence of the orthogonality between trion and exciton states without resorting to the polarization of the electron gas. 

The dependence of the trion and exciton energies on the order of electrons is further explored by repeating the calculations in square and honeycomb 2D lattices. The supplemental material provides detailed information of their exciton band structures \cite{supp}. Figure~\ref{fig:TrionExc}(c) shows the  energy dependence of the trion ($X^-$) and exciton  ($X^0$) states on electron density in the three studied lattices. The dashed lines correspond to $\Gamma$-point energy gaps in the nearly-free exciton model of each lattice type,   
\begin{equation}
E^\square_{\Gamma} =  \frac{\hbar^2}{M} 2\pi^2 n \quad \text{and} \quad E^{\hexagon}_{\Gamma} =  \tfrac{1}{2}E^\triangle_{\Gamma} = \frac{\hbar^2}{M}  \frac{2\pi^2}{\sqrt{3}}n\,\,. 
\end{equation}

Figure~\ref{fig:TrionExc}(c) shows that the trion energy weakly depends on the lattice type, whereas the exciton energy shows clear dependence. The reason is that delocalized exciton states can better `sense' the lattice structure at a larger range from the lattice site. To understand the dependence on lattice types, we write the smallest distance between two neighboring electrons in each lattice: $r_s^\triangle  \simeq 1.07\,\,  n^{-1/2}$, $r_s^\square  = n^{-1/2}$, and $r_s^{\hexagon}  \simeq 0.877 \,\,n^{-1/2}$. This parameter is a measure of how close electrons are packed, indicating that for a certain change in electron density, the resulting change in $r_s$ is largest in the triangular lattice. As such, the energy blueshift of the exciton in this lattice has the strongest dependence on density. 

The dependence of the exciton energy blueshift on the lattice type is an important result. It suggests that the blueshift is stronger when the lattice tends to structures with higher symmetry. Such a change in energy blueshift of the exciton can be used as a precursor to Wigner crystallization \cite{Smolenski_Nat21}. For example, the energy blueshift should become weaker when the temperature increases due to reduced order of the lattice. The Wigner crystal should also become disordered by continuing to increase the electron density (when $r_s$ is small enough) \cite{Cassella_PRL23,Giuliani_Vignale_Book}. The broadening of the exciton peak when the charge density increases or at elevated temperatures can be modeled by spatial fluctuations of $r_s$. Finally, we mention that regardless of order, the energy blueshift effect is expected to be relatively strong when the electron of the exciton has different quantum numbers than those of the semiconductor electrons (spin and/or valley).  Otherwise, Pauli exclusion helps the exciton to avoid indistinguishable electrons  \cite{VanTuan_PRL22, Tiene_PRB22}, leading to weaker energy blueshift due to exchange interaction. 

In conclusion, we have presented a band theory to study the behavior of excitons and trion states in electrostatically-doped semiconductors. We have explained how energies of these states shift as a function of electron density, and provided important guidelines to detect the elusive Wigner crystal if the electrons of the host semiconductor are ordered. Ramifications of this study can be used to emulate various lattices in which excitons propagate in the same manner that electrons propagate in atomic crystals or light in photonic crystals. Beyond Wigner crystals, such lattices can emerge in Moir\'{e} heterostructures or atom-decorated monolayer semiconductors. In the latter case, advanced fabrication techniques can engineer artificial lattices with various symmetries through  controlled placement of atoms or molecules in or adjacent to 2D semiconductors. Such fabricated lattices can be used as far-infrared light detectors, as a means to enhance the functionality of polariton cavities, or as a platform to further investigate excitonic band structures. 


\acknowledgments{We thank Mikhail Glazov for fruitful discussions and for bringing to our attention the work of Voronov and Ivchenko  \cite{Voronov_PSS03}.   D. V. T. is supported by the Department of Energy, Basic Energy Sciences, Division of Materials Sciences and Engineering under Award No. DE-SC0014349. H. D. is supported by the Office of Naval Research, under Award No. N000142112448. }



\begin{thebibliography}{99}

\bibitem{HaugBook} H. Haug and S. W. Koch, \textit{Quantum theory of the optical and electronic properties of semiconductors}, 3rd ed. (World Scientific, Singapore, 1994).

\bibitem{Combescot_Book} M. Combescot and S.-Y. Shiau, \textit{Excitons and Cooper pairs: two composite bosons in many-body physics}, (Oxford University Press, Singapore, 2016).

\bibitem{Wang_RMP18}  G. Wang, A. Chernikov, M. M. Glazov, T. F. Heinz, X. Marie, T. Amand, and B. Urbaszek, Colloquium: Excitons in atomically thin transition metal dichalcogenides, Rev. Mod. Phys. \textbf{90}, 021001 (2018).

\bibitem{Mahan_PR67b} G. D. Mahan, Excitons in degenerate semiconductors, Phys. Rev. \textbf{153}, 882 (1967).

\bibitem{Skolnick_PRL87} M. S. Skolnick , J. M. Rorison, K. J. Nash, D. J. Mowbray, P. R. Tapster, S. J. Bass, and A. D. Pitt, Observation of a many-body edge singularity in quantum-well luminescence spectra, Phys. Rev. Lett. \textbf{58}, 2130 (1987).


\bibitem{VanTuan_PRL22}  D. V. Tuan, S.-F. Shi, X. Xu, S. A. Crooker, and H. Dery,  Six-body and eight-body exciton states in monolayer WSe$_2$, Phys. Rev. Lett. \textbf{129}, 076801 (2022).

\bibitem{VanTuan_PRB22}  D. V. Tuan and H. Dery,  Composite excitonic states in doped semiconductors, Phys. Rev. B \textbf{106}, L081301 (2022).



\bibitem{Scharf_PRL17}
B. Scharf, G. Xu, A. Matos-Abiague, and I. \v{Z}uti\'c, Magnetic proximity effects in transition-metal dichalcogenides: Converting excitons, Phys. Rev. Lett. \textbf{119}, 127403 (2017).

\bibitem{Choi_NatMater22}
J. Choi, C. Lane, J.-X. Zhu, and S. A. Crooker,   Asymmetric magnetic proximity interactions in MoSe$_2$/CrBr$_3$ van der Waals heterostructures, Nat. Mater. (2022). \url{https://doi.org/10.1038/s41563-022-01424-w}


\bibitem{Sidler_NatPhys17} M. Sidler, P. Back, O. Cotlet, A. Srivastava, T. Fink, M. Kroner, E. Demler, and A. Imamoglu, Fermi polaron-polaritons in charge-tunable atomically thin semiconductors, Nat. Phys. \textbf{13}, 255 (2017).

\bibitem{Ravets_PRL18} S. Ravets, P. Kn\"{u}ppel, S. Faelt, O. Cotlet, M. Kroner, W. Wegscheider, and A. Imamoglu, Polaron polaritons in the integer and fractional quantum Hall regimes,  Phys. Rev. Lett. \textbf{120}, 057401 (2018).


\bibitem{Smolenski_Nat21}
T. Smolenski, P. E. Dolgirev, C. Kuhlenkamp, A. Popert, Y. Shimazaki, P. Back, X. Lu, M. Kroner, K. Watanabe, T. Taniguchi, I. Esterlis, E. Demler, and  A. Imamoglu, Signatures of Wigner crystal of electrons in a monolayer semiconductor, Nature \textbf{595}, 53 (2021).

\bibitem{Zhou_Nat21}
Y. Zhou, J. Sung, E. Brutschea, I. Esterlis, Y. Wang, G. Scuri, R. J. Gelly, H. Heo, T. Taniguchi, K. Watanabe, G. Zarand, M. D. Lukin, P. Kim, E. Demler, and H. Park, Bilayer Wigner crystals in a transition metal dichalcogenide heterostructure, Nature \textbf{595}, 48 (2021).
 
 
\bibitem{Regan_Nat20} 
E. C. Regan, D. Wang, C. Jin, M. I. Bakti Utama, B. Gao, X. Wei, S. Zhao, W. Zhao, Z. Zhang, K. Yumigeta, M. Blei, J. D. Carlstr\"{o}m, K. Watanabe, T. Taniguchi, S. Tongay, M. Crommie, A. Zettl, and F. Wang, Mott and generalized Wigner crystal states in WSe$_2$/WS$_2$ moir\'{e} superlattices, Nature \textbf{579}, 359 (2020).

 \bibitem{Xu_Nat20} 
Y. Xu, S. Liu, D. A. Rhodes, K. Watanabe, T. Taniguchi, J. Hone, V. Elser, K. F. Mak, and J. Shan, Correlated insulating states at fractional fillings of moir\'{e} superlattices, Nature \textbf{587}, 214 (2020).



\bibitem{SchmittRink_PRB86} S. Schmitt-Rink, C. Ell, and H. Haug, Many-body effects in the absorption, gain and luminescence spectra of semiconductor quantum-well structures, Phys. Rev. B \textbf{33}, 1183 (1986).

\bibitem{VanTuan_PRB19} D. V. Tuan, B. Scharf, Z. Wang, J. Shan, K. F. Mak, I. \v{Z}uti\'c, and H. Dery, Probing many-body interactions in monolayer transition-metal dichalcogenides, Phys. Rev. B \textbf{99}, 085301 (2019). 


\bibitem{Haug_PQE} H. Haug and S. Schmitt-Rink, Electron theory of the optical
proporties of laser excited semiconductors, Prog. Quant. Electr. \textbf{9}, 3 (1984).

\bibitem{Hawrylak_PRB91} P. Hawrylak, Optical properties of a two-dimensional electron gas: evolution of spectra from excitons to fermi-edge singularities, Phys. Rev. B  \textbf{44}, 3821 (1991).

\bibitem{Bronold_PRB00} F. X. Bronold, Absorption spectrum of a weakly $n$-doped semiconductor quantum well, Phys. Rev. B. \textbf{61}, 12620 (2000).

\bibitem{Suris_PSSb01} R. A. Suris, V. P. Kochereshko, G. V. Astakhov, D. R. Yakovlev, W. Ossau, J. Nurnberger, W. Faschinger, G. Landwehr, T. Wojtowicz, G. Karczewski, and J. Kossut, Excitons and trions modified by interaction with a two-dimensional electron gas, Phys. Stat. Sol. (b) \textbf{227}, 343
(2001).

\bibitem{Esser_PSSb01} A. Esser, R. Zimmermann, and E. Runge, Theory of trion spectra in semiconductor nanostructures, Phys. Stat. Sol. (b)\textbf{227}, 317 (2001). 

\bibitem{VanTuan_PRX17} D. Van Tuan, B. Scharf, I. \v{Z}uti\'c, and H. Dery, Marrying excitons and plasmons in monolayer transition-metal dichalcogenides, Phys. Rev. X \textbf{7}, 041040 (2017).

\bibitem{Efimkin_PRB17}  D. K. Efimkin and A. H. MacDonald,  Many-body theory of trion absorption features in two-dimensional semiconductors, Phys. Rev. B \textbf{95}, 035417 (2017).

\bibitem{Chang_PRB19} Y.-W. Chang and D. R. Reichman, Many-body theory of optical absorption in doped two-dimensional semiconductors, Phys. Rev. B 99, 125421 (2019).

\bibitem{Glazov_JCP20} M. M. Glazov, Optical properties of charged excitons in two-dimensional semiconductors, J. Chem. Phys. \textbf{153}, 034703 (2020).

\bibitem{Rana_PRB20} F. Rana, O. Koksal, and C. Manolatou, Many-body theory of the optical conductivity of excitons and trions in two-dimensional materials, Phys. Rev. B \textbf{102}, 085304 (2020).

\bibitem{Imamoglu_CR21} A. Imamoglu, O. Cotlet, R. Schmidt, Exciton-polarons in two-dimensional semiconductors and the Tavis-Cummings model, Comptes Rendus. Physique \textbf{22}, 89  (2021). 

\bibitem{Wang_NanoLett17} Z. Wang, L. Zhao, K. F. Mak, and J. Shan, Probing the spin-polarized electronic band structure in monolayer transition metal dichalcogenides by optical spectroscopy, Nano Lett. \textbf{17}, 740 (2017).

\bibitem{Courtade_PRB17} E. Courtade, M. Semina, M. Manca, M. M. Glazov, C. Robert, F. Cadiz, G. Wang, T. Taniguchi, K. Watanabe, M. Pierre, W. Escoffier, E. L. Ivchenko, P. Renucci, X. Marie, T. Amand, and B. Urbaszek, Charged excitons in monolayer WSe$_2$: experiment and theory, Phys. Rev. B \textbf{96}, 085302 (2017). 

\bibitem{Smolenski_PRL19}   T. Smole\'{n}ski, O. Cotlet, A. Popert, P. Back, Y. Shimazaki, P. Kn\"{u}ppel, N. Dietler, T. Taniguchi, K. Watanabe, M. Kroner, and A. Imamoglu,  Interaction-induced Shubnikov–de Haas oscillations in optical conductivity of monolayer  MoSe$_2$, Phys. Rev. Lett. \textbf{123}, 097403 (2019).

\bibitem{Liu_PRL20} E. Liu, J. van Baren, T. Taniguchi, K. Watanabe, Y.-C. Chang, and C. H. Lui, Landau-quantized excitonic absorption and luminescence in a monolayer valley semiconductor, Phys. Rev. Lett. \textbf{124}, 097401 (2020).

\bibitem{Wang_PRX20} T. Wang, Z. Li, Z. Lu, Y. Li, S. Miao, Z. Lian, Y. Meng, M. Blei, T. Taniguchi, K. Watanabe, S. Tongay, W. Yao, D. Smirnov, C. Zhang, and S.-F. Shi, Observation of quantized exciton energies in monolayer WSe$_2$ under a strong magnetic field, Phys. Rev. X \textbf{10}, 021024 (2020).

\bibitem{Li_NanoLett22} J. Li, M. Goryca, J. Choi, X. Xu, S. A. Crooker, Many-body exciton and intervalley correlations in heavily electron-doped WSe$_2$ monolayers, Nano Lett. \textbf{22}, 426 (2022). 

\bibitem{Liu_NatComm21} E. Liu, J. van Baren, Z. Lu, T. Taniguchi, K. Watanabe, D. Smirnov, Y.-C. Chang, and C.-H. Lui, Exciton-polaron Rydberg states in monolayer MoSe$_2$ and WSe$_2$, Nat. Commun. \textbf{12}, 6131 (2021). 

\bibitem{VanTuan_PRB2012} D. V. Tuan, A. Kumar, S. Roche, F. Ortmann, M. F. Thorpe, and P. Ordejon, Insulating behavior of an amorphous graphene membrane, Phys. Rev. B \textbf{86}, 121408(R) (2012).


\bibitem{Wigner_PR34}
E. Wigner, On the interaction of electrons in metals, Phys. Rev. \textbf{46}, 1002 (1934). 


\bibitem{Voronov_PSS03} M. M. Voronov and E. L. Ivchenko, Resonance reflection of light from two-dimensional superlattice structures, Phys. Sol. State \textbf{45}, 176 (2003). Translated from Fizika Tverdogo Tela,  \textbf{45}, 168 (2003). 

\bibitem{Shimazaki_PRX17} Y. Shimazaki, C. Kuhlenkamp, I. Schwartz, T. Smole\'{n}ski, K. Watanabe, T. Taniguchi, M. Kroner, R. Schmidt, M. Knap, and A. Imamoglu, Optical signatures of periodic charge distribution in a Mott-like correlated insulator state,  Phys. Rev. X \textbf{11}, 021027 (2021).

\bibitem{footnote} Complete distinguishability between the electron of the exciton and all other conduction-band electrons is naturally achieved in WSe$_2$ monolayers \cite{VanTuan_PRL22}. In MoSe$_2$ or hole-doped WSe$_2$ monolayers, such complete distinguishability is achieved by applying a strong magnetic field to populate a single spin-polarized valley and create excitons in the opposite valley \cite{Wang_PRX20,Liu_PRL20,Li_NanoLett22}, or by valley pumping in tungsten-based monolayers  \cite{Robert_NatComm21}.

\bibitem{Robert_NatComm21} C. Robert, S. Park, F. Cadiz, L. Lombez, L. Ren, H. Tornatzky, A. Rowe, D. Paget, F. Sirotti, M. Yang, D. V. Tuan, T. Taniguchi, B. Urbaszek, K. Watanabe, T. Amand, H. Dery, and X. Marie, Spin/valley pumping of resident electrons in WSe$_2$ and WS$_2$ monolayers, Nat. Commun. \textbf{12}, 5455 (2021).

\bibitem{Chelikowsky_PRB76}
J. R. Chelikowsky and M. L. Cohen, Nonlocal pseudopotential calculations for the electronic structure of eleven diamond and zinc-blende semiconductors, Phys. Rev. B. \textbf{14} 556 (1976).

\bibitem{Cardona_book} P. Y. Yu and M. Cardona, \textit{Fundamentals of semiconductors}, 3rd ed. (Springer, Berlin, 2005).





\bibitem{Yang_PRB22} M. Yang ,L. Ren, C. Robert, D. V. Tuan , L. Lombez,
B. Urbaszek, X. Marie, and H. Dery, Relaxation and darkening of excitonic complexes in electrostatically doped monolayer WSe$_2$: Roles of exciton-electron and trion-electron interactions,  Phys. Rev. B \textbf{105}, 085302 (2022). 








\bibitem{supp} See Supplemental Material at http://link.aps.org/supplemental... for details of the band structures and symmetry analysis of square and honeycomb 2D lattices.


\bibitem{Giuliani_Vignale_Book} G. Giuliani and G. Vignale, \textit{Quantum theory of the electron liquid} (Cambridge University Press, Cambridge, 2005).

\bibitem{Cassella_PRL23} G. Cassella, H. Sutterud, S. Azadi, N. D. Drummond, D. Pfau, J. S. Spencer, and W. M. C. Foulkes, Discovering quantum phase transitions with Fermionic neural networks, Phys. Rev. Lett. \textbf{130}, 036401 (2023).


\bibitem{Tiene_PRB22} A. Tiene, J. Levinsen, J. Keeling, M. M. Parish, and F. M. Marchetti, Effect of fermion indistinguishability on optical absorption of doped two-dimensional semiconductors,  Phys. Rev. B \textbf{105}, 125404 (2022).






\end{thebibliography}
\end{document}